\documentclass[11pt]{article}

\usepackage{graphicx}
\usepackage{float}
\usepackage{amssymb}
\usepackage{chet}
\usepackage{subfig}
\usepackage{tabularx}
\usepackage{dsfont}
\usepackage{multirow}

\def\bZ{\mathbb{Z}}

\newcommand{\lsp}{\hspace{1pt}}
\newcommand{\llsp}{\hspace{0.5pt}}

\DeclareMathOperator{\Tr}{Tr}

\def\be{\begin{equation}}
\def\ee{\end{equation}}
\def\bea{\begin{eqnarray}}
\def\eea{\end{eqnarray}}
\def\ie{\begin{equation}\begin{aligned}}
\def\fe{\end{aligned}\end{equation}}
\def\ge{\geqslant}
\def\le{\leqslant}
\def\geq{\geqslant}
\def\leq{\leqslant}

\definecolor{darkblue}{rgb}{0.1,0.1,0.7}
\hypersetup{colorlinks,
           linkcolor={darkblue},
           citecolor={darkblue},
           urlcolor={darkblue}
}

\preprint{CALT-TH-2019-034\\
LA-UR-19-30323}

\title{Bounds on Triangle Anomalies in (3+1)$d$}

\author{Ying-Hsuan Lin,$\!^a$ David Meltzer,$\!^a$ Shu-Heng Shao,$\!^b$ and Andreas Stergiou$^c$}

\affiliation{${}^a$Walter Burke Institute for Theoretical Physics,
  California Institute of Technology,\\\vspace{-3pt}
  Pasadena, CA 91125, USA\\
  ${}^b$School of Natural Sciences, Institute for Advanced Study,
  Princeton, NJ 08540, USA\\
  ${}^c$Theoretical Division, MS B285, Los Alamos National Laboratory, Los
  Alamos, NM 87545, USA
  \emails{yhlin@caltech.edu, dmeltzer@caltech.edu, shao@ias.edu,
  andreas@lanl.gov}
}

\abstract{{\it How many charged degrees of freedom are necessary to accommodate a certain amount of 't Hooft anomaly?} Using the conformal bootstrap for the four-point function of flavor current multiplets, we show that in all (3+1)$d$ superconformal field theories the 't Hooft anomaly of a continuous flavor symmetry is bounded from above by
the $3/2$ power of the current two-point function coefficient,  which can be thought of as a measure for the amount of charged degrees of freedom. We check our bounds against free fields and SQCD in the conformal window. }

\date{September 2019}

\begin{document}

\maketitle

\toc

\section{Introduction}

The modern landscape of quantum field theory has been enriched by surprising dualities. Key to these discoveries are tools for investigating the seemingly untamable nonperturbative phenomena.  One of the most prominent tools is perhaps 't Hooft anomalies, which characterize the obstruction to gauging a global symmetry.
't Hooft anomalies are robust under symmetry-preserving deformations, and are therefore under control even in the nonperturbative regime of quantum field theory. For instance, the matching of 't Hooft anomalies provided strong evidence for Seiberg's proposed dualities for (3+1)$d$ supersymmetric quantum chromodynamics (SQCD) \cite{Seiberg:1994pq}.

A new tool that has emerged in recent years is the conformal
bootstrap~\cite{Rattazzi:2008pe} (see \cite{Poland:2018epd, Chester:2019wfx} for recent
reviews), which exploits unitarity and conformal symmetry in the gapless phase of quantum field theory to directly {\it solve} for nonperturbative observables, and does not rely on the fixed point's proximity (in the sense of renormalization group flow) to free theories.  This bootstrap revolution has had particular success with the (2+1)$d$ Ising CFT~\cite{ElShowk:2012ht, El-Showk:2014dwa, Kos:2014bka} and has corroborated and further improved upon the best prior Monte Carlo studies~\cite{Kos:2016ysd}.

In this work, we unite these two tools and study universal bootstrap bounds on the 't Hooft anomalies of global symmetries.  Using the modular bootstrap, the implications of  the 't Hooft anomalies (for both the $\bZ_2$ and the $U(1)$ global symmetry) for (1+1)$d$ conformal field theory have been studied in \cite{Lin:2019kpn}.
Here we will study (3+1)$d$ conformal field theory with continuous global symmetries.  For simplicity, we limit ourselves to $U(1)$ global symmetry in this section, while the nonabelian cases will be discussed in the later sections.  Let the Noether current associated with the $U(1)$ global symmetry be $j_\mu(x)$ and let $A_\mu$ be its one-form background gauge field.  The 't Hooft anomaly is captured by the nonconservation of $j_\mu$ in the presence of a nontrivial $A_\mu$:
\ie
\label{nonconservation}
d \star j =  {i \over 4\pi^2 } {K\over 3!} F\wedge F\,,
\fe
where $F=dA$ is the field strength.  With the understanding that $j$ and $A$ are properly normalized, this anomaly is  completely  encoded in the coefficient $K$. In (3+1)$d$ Lagrangian theories, the quantity $K$ can be computed by a triangle diagram where all three legs are taken to be the background gauge field $A_\mu$ for the $U(1)$ global symmetry, hence this anomaly is commonly referred to as the {\it triangle} anomaly.\footnote{The 't Hooft anomaly is to be distinguished from the Adler--Bell--Jackiw (ABJ) anomaly \cite{Adler:1969gk, Bell:1969ts}, where the relevant triangle diagram involves one background gauge field for the (classical) $U(1)$ global symmetry (the axial ``symmetry"), and two dynamical gauge fields for a $U(1)$ gauge symmetry.  The classical  axial ``symmetry" fails to be a true global symmetry of the quantum field theory because of the ABJ anomaly.  By contrast, a global symmetry with 't Hooft anomaly is still a true global symmetry in a consistent quantum field theory, but there is an obstruction to gauging it.
}
Nonetheless, it is a universal quantity in (3+1)$d$ quantum field theory, well-defined even at strong coupling or in the absence of a Lagrangian formulation.

We ask: {\it How many charged degrees of freedom are necessary to accommodate a certain amount of the triangle anomaly?}

The question above can be motivated by first examining free massless Weyl fermions. Consider a $U(1)$ global symmetry such that the $i$-th Weyl fermion has charge $q_i\in \bZ$. The only triangle diagrams that contribute to $K$ are those with massless charged fields running in the loop, and we have \ie K = \sum_{i \in L} q_i^3 - \sum_{i \in R} q_i^3, \fe where $L$ and $R$ are the free Weyl fermions of left and right chirality, respectively. The two-point function of the  Noether current for the $U(1)$ global symmetry takes the form
\ie
\label{jjform}
\langle  j_\mu(x) j_\nu(0)\rangle = 3\tau  {I_{\mu\nu}(x) \over 4\pi^4
r^6}\,,
\fe
where $I_{\mu\nu}(x) \equiv \delta_{\mu\nu} - {2 x_\mu x_\nu \over r^2}$ and $r^2 = x^\mu x_\mu$. Since the Noether current is just a sum of free field bilinears, Wick contractions determine the current two-point function coefficient $\tau$:
\ie
\tau =\tfrac23 \sum_{i \in L \cup R} q_i^2\,,
\fe
which can be taken as a quantifier for the amount of ``charged degrees of freedom''.\footnote{The current two-point function is just one crude estimate for the charged degrees of freedom.  In particular, it is not always a monotonically decreasing function in renormalization group flows \cite{Anselmi:1997am}.}
Clearly, to have $K$ amount of triangle anomaly, we need at least $\tau = {2\over3}|K|^{2/3}$ worth of charged degrees of freedom, {\it i.e.}\ the quantity $|K|/\tau^{3/2}$ has an upper bound in free field theory.\footnote{This bound in free field theory follows from the decreasing nature of the $p$-norm in $p$, or from this elementary proof:
\ie
(\tfrac32\tau)^3 - K^2 = (\sum_i q_i^2)^3 - (\sum_i q_i^3)^2 = \sum_{\substack{i,j,k \\ \text{distinct}}} q_i^2 q_j^2 q_k^2 + 2 \sum_{i \neq j} q_i^4 q_j^2 + \sum_{i > j} (q_i^4 q_j^2 + q_i^2 q_j^4 - 2 q_i^3 q_j^3) \geq 0 \, ,
\fe
where $q_i^4 q_j^2 + q_i^2 q_j^4 \geq 2 |q_i|^3 |q_j|^3 \geq 2 q_i^3 q_j^3$ by the AM-GM inequality.
}

This intuition from free theory leads us to expect that similar bounds
exist for general quantum field theory, once a suitable notion of charged
degrees of freedom is given. In conformal field theory, the current
two-point function also takes the form \eqref{jjform}, so we immediately
have a suitable notion of the amount of charged degrees of freedom. The
main goal of this paper is to explore bounds on $|K|/\tau^{3/2}$.

In practice, there are technical difficulties in obtaining bounds on the triangle anomaly for general (3+1)$d$ conformal field theory, due to the complexity of bootstrapping spinning correlators~\cite{Dymarsky:2017xzb, Dymarsky:2017yzx, Cuomo:2017wme, Karateev:2019pvw}.\foot{The bootstrap of conserved spin-one abelian currents in (3+1)$d$ is underway~\cite{KKMSVprep}.}  In this paper, we take a shortcut by considering bounds on the triangle anomaly of {\it supersymmetric} theories.  Supersymmetry relates flavor current correlators to scalar correlators lying in the same supermultiplet, thereby greatly reducing the complexity of the problem.\footnote{Note that the $R$-symmetry spin-one current is in the stress tensor multiplet and is itself the superconformal primary.  Since it is in a different kind of multiplet from the flavor current, bootstrapping its correlators  lies outside the scope of this work.}  Supersymmetry also relates the two-point function coefficient $\tau$ to the mixed flavor-$R$ anomaly, which is invariant under renormalization group flows.\footnote{Without supersymmetry, the  two-point function coefficient $\tau$ is only related to the mixed flavor-conformal anomaly, which is generally not an RG invariant.
}

The current two-point function coefficient $\tau$ has been bounded in (3+1)$d$ superconformal field theories by bootstrapping the four-point functions of charged scalar operators  in \cite{Poland:2010wg, Poland:2011ey}.  The superconformal bootstrap of flavor current multiplet correlators in (3+1)$d$ supersymmetric theories has also been previously studied in \cite{Berkooz:2014yda,Li:2017ddj}.  In this paper, we focus instead on bounding the triangle anomaly.  In particular, our results present interesting comparisons with the conformal window of SQCD.

The remainder of this paper is organized as follows. Section~\ref{sec:Warm-up} illustrates the idea of bounding the 't Hooft anomaly per charged degree of freedom in the simple (1+1)$d$ case. Section~\ref{sec:PertAnomalies} reviews perturbative 't Hooft anomalies in (3+1)$d$ quantum field theory, and then further specializes to conformal field theory and superconformal field theory. Section~\ref{sec:Bootstrap} sets up the bootstrap of current multiplet scalars in superconformal field theory, and presents the resulting numerical bounds on $|K|/\tau^{3/2}$. A comparison of the $SU(N)$ bounds and the conformal window of SQCD is also given. Section~\ref{sec:Discussion} summarizes the results, and comments on holography and future directions. Finally, Appendix~\ref{app:tau} presents the improved bounds on $\tau$.

\section{Warm-up: Bounds on the \texorpdfstring{$\boldsymbol{U(1)}$}{U(1)} Anomaly in (1+1)\texorpdfstring{$\boldsymbol{d}$}{d}}
\label{sec:Warm-up}

We start  with a warm-up question: Is there an upper bound on the 't Hooft anomaly for a $U(1)$ global symmetry in (1+1)$d$?  We will follow the exposition in Section 6 of \cite{Lin:2019kpn}.

Let $J_\mu(z,\bar z)$ be the Noether  current associated to a compact $U(1)$ global symmetry satisfying the conservation equation $\partial^\mu J_\mu(x)=0$. We will denote the holomorphic and the antiholomorphic components of the current as $J\equiv J_z$ and $\bar J\equiv J_{\bar z}$, respectively.  Note that we do \textit{not} restrict ourselves to holomorphic (nor antiholomorphic) $U(1)$,  {\it i.e.}\ we do not assume $\bar J=0$ ($J=0$).

In any compact unitary (1+1)$d$ CFT, unitarity further implies that $\partial\bar J= 0$ and $\bar\partial J=0$, so each of them is separately a $\mathfrak{u}(1)$ Lie algebra generator.  Globally, however, the holomorphic current $J(z)$ generically does not  generate a compact $U(1)$ group, but rather an $\mathbb{R}$. The same is true for the anti-holomorphic current $\bar J$. For example, the $c=1$ compact boson at a generic radius has $U(1)_n\times U(1)_w$ global symmetry, but neither of the two $U(1)$'s is holomorphic or antiholomorphic.  Indeed, the holomorphic $\mathfrak{u}(1)$ charge $n/R+wR$ of a local operator $\exp\left[ i (n/R+wR)X_L(z) + i (n/R-wR)X_R(\bar z)\right]$  is irrational at a generic radius $R$.

In higher dimensions, there is a unique tensor structure for the two-point function of a conserved spin-one current.  In (1+1)$d$, however, there are two independent structures:
\begin{align}
\langle J(z) J(0)\rangle ={k \over z^2}\,,~~~~~\langle\bar J(\bar z) \bar J(0)\rangle = {\bar k\over \bar z^2}\,.
\end{align}
Note that the levels $k\ge0$ and $\bar k\ge0$ are physically meaningful and cannot be scaled away if we assume that our symmetry is globally a $U(1)$ acting faithfully on all local operators.  On the one hand, we define $\tau$ in (1+1)$d$  as the average of the two levels:
\ie \tau \equiv {k+\bar k\over2}\,. \fe
On the other hand, the 't Hooft anomaly of the $U(1)$ global symmetry in a bosonic (1+1)$d$ CFT is
\begin{align}
K\equiv {k-\bar k\over2} \in \bZ\,.
\end{align}
Therefore, we see that the 't Hooft anomaly $K$ is trivially bounded from above by the current two-point function coefficient $\tau$:
\ie
|K|\le \tau \,.
\fe
In the rest of the paper we will use the conformal bootstrap in (3+1)$d$ to derive an analogous upper bound on the 't Hooft anomaly of a global symmetry by a power of the current two-point function coefficient.

\section{Perturbative 't Hooft Anomalies in
(3+1)\texorpdfstring{$\boldsymbol{d}$}{d}}
\label{sec:PertAnomalies}

Consider a (3+1)$d$ conformal field theory with a $U(1)$ global symmetry, possibly with a 't Hooft anomaly.  Let the associated conserved Noether current be $j_\mu(x)$.  To detect the anomaly,  we  couple the theory to a one-form background gauge field $A_\mu$ via the coupling $\int d^4x \lsp j_\mu A^\mu$.  The hallmark of the 't Hooft anomaly is that the partition function $Z[A]$ in the presence of this background is not invariant under the background gauge transformation $  A_\mu \to A_\mu +\partial_\mu\Lambda$, but changes by a phase:\footnote{Here we ignore the possible phases from the mixed gauge-gravitational anomaly. }
\ie\label{Z}
 Z[A+d \Lambda] =
  \exp\left[- {i\over 4\pi^2}  {K \over 3!} \int \Lambda F\wedge F \right]  \, Z[A] \,,
\fe
where the anomaly coefficient $K\in \bZ$ is normalized to be 1 for a left-handed Weyl fermion with plus one  $U(1)$ charge.  While the $U(1)$ is a perfectly healthy global symmetry with a conserved Noether current, the anomalous phase in \eqref{Z} signals the obstruction to gauging the global symmetry $U(1)$.  This is the conventional perturbative 't Hooft anomaly captured by the one-loop triangle Feynman diagram in a Lagrangian theory.

The anomalous transformation \eqref{Z} implies that the current is not conserved in the presence of a nontrivial background field, with the nonconservation given in \eqref{nonconservation}.  By taking the functional derivative with respect to $A_\mu$, we see that the anomaly coefficient $K$ enters into the contact term of the three-point
function $\langle \partial^\mu j_\mu(x) j_\nu(y) j_\rho(z)\rangle$.  Upon integration,  $K$ becomes the coefficient of the  parity-odd structure in the three-point function $\langle j_\mu(x) j_\nu(y) j_\rho(z)\rangle$ at separated points \cite{Schreier:1971um}.

The normalization of the current $j_\mu(x)$ is fixed by the Ward identity,
\ie
\partial^\mu j_\mu(x) {\cal O}(y) =  iq\lsp\delta^{(4)}(x-y){\cal O}(y)\,,
\fe
where ${\cal O}$ is a local operator with $U(1)$ charge $q\in \bZ$. Therefore, the overall  coefficient $\tau$ of the two-point function for  $j_\mu(x)$, which takes the form of \eqref{jjform} in any CFT, is physically meaningful.  In a free theory of $N$ Weyl fermions with $U(1)$ charges $q_i$, $\tau \sim \sum_i q_i^2$, so roughly speaking $\tau$ measures the amount of charged degrees of freedom.

In the conformal bootstrap we normalize external operators, {\it i.e.}\ the ones in the four-point function under consideration, to have unit two-point function coefficients.  The three-point function of the normalized current $\hat{\jmath} _\mu = {2\pi^2\over \sqrt{3\tau} } j_\mu$ is
\ie
\langle \hat{\jmath}_\mu (x)\hat{\jmath}_\nu(y) \hat{\jmath}_\rho(z)\rangle = {  K \over\tau^{3/2}} D_{\mu\nu\rho}(x,y,z)\,,
\fe
where $D_{\mu\nu\rho}(x,y,z)$ is a parity-odd structure that is fixed by conformal symmetry.  The main point of the current paper is to use the conformal bootstrap of the current four-point function to place an upper bound on the three-point function coefficient ${ | K |/\tau^{3/2}}$.

In a nonsupersymmetric theory, the constraints we are after would require bootstrapping the four-point function of the spin-one conserved current $j_\mu(x)$.  However, if we assume ${\cal N}=1$ supersymmetry, then the conserved, flavor current resides in a multiplet whose zero component is a real scalar, $J(x)$.  Furthermore, the associated superconformal blocks for the four-point function $\langle J(x_1) J(x_2) J(x_3) J(x_4)\rangle$ are known \cite{Fortin:2011nq, Fitzpatrick:2014oza, Khandker:2014mpa}.  Due to these simplifications, we will restrict ourselves to (3+1)$d$ ${\cal N}=1$ superconformal field theory, but our arguments can be generalized to (3+1)$d$ nonsupersymmetric conformal field theory. Note that if we further impose ${\cal N}=2$ or higher supersymmetry, then the theory is necessarily nonchiral, and the 't Hooft anomaly coefficient $K$ is zero.

The OPE between two $J(x)$'s is \cite{Fortin:2011nq}\footnote{Our anomaly coefficient  $K$ is related to $\kappa$ and $k$ in \cite{Berkooz:2014yda} as follows.  When the symmetry group is $U(1)$, $K=\kappa/8$. When the symmetry group is $SU(N)$, $K = k/4$. }
\ie
J(x) J(0)  = {\tau \over 16\pi^4 r^4}\mathds{1}  +{ K \over \tau } {J(0)\over 2 \pi^2r^2} +\cdots\,.
\fe
Here the normalization of $J(x)$ relative to $j_\mu(x)$ is fixed by $\langle J(x) J(0)\rangle = {\tau /16\pi^4r^4}$.  In a free chiral multiplet, $J(x)$ for the $U(1)$ flavor symmetry is given by $J(x) = \phi^\dagger (x) \phi(x)$, where $\phi(x)$ is a complex  free scalar normalized as $\langle \phi^\dagger(x) \phi(0)\rangle= 1/4\pi^2 r^2$.  For a charge $+1$ free chiral multiplet, $K=1$ and $\tau=1$, hence $K/\tau^{3/2}=1$.

We will also consider (3+1)$d$ SCFTs with $SU(N)$ global symmetry. The $J^a\times J^b$ OPE is
\ie
J^a(x) J^b(0)  = {\tau \,  \delta^{ab}\over 16\pi^4 r^4}\mathds{1}  +{ K \over \tau }
d^{\llsp abc} {J^c(0)\over 4 \pi^2r^2} + f^{abc} { x^\mu j_\mu^c \over 24\pi^2 r^2} +\cdots\,,
\label{JJOPE}
\fe
where $d^{abc}  = \Tr_\square (\{ T^a ,T^b \} T^c)$ and $f^{abc} = -i \Tr_\square ([T^a,T^b]T^c)$.  Here the $\square$ means that the trace is taken in the fundamental representation, in which we normalize the $SU(N)$ generator to be $\Tr_\square(T^aT^b)  = \delta^{ab}$.  The triangle anomaly only exists if $N\ge3$, since $d^{\llsp abc}=0$ for $SU(2)$.

In the free theory of $N$ chiral multiplets, $\phi^i$ ($i=1,\ldots ,N$), the $SU(N)$ flavor symmetry current is  $J^a= \phi^\dagger_i (T^a)^i_{~j} \phi^j$, with the scalar normalized as $\langle \phi^\dagger_i (x) \phi^j(0)\rangle = \delta_i^{\; j}/4\pi^2 r^2$.  The anomaly term in the $J^a\times J^b$ OPE is
\ie
{1\over 4\pi^2 r^2} \phi^\dagger_i (x) (T^a)^i_{~j}(T^b)^j_{~\ell} \phi^\ell(0)
+{1\over 4\pi^2 r^2} \phi^\dagger_i (0) (T^b)^i_{~j}(T^a)^j_{~\ell}
\phi^\ell(x) = {1\over 4\pi^2r^2} d^{\llsp abc} J^c(0)+\cdots\,.
\fe
Hence, for the free theory of $SU(N)$ fundamental scalars, we have $K/\tau^{3/2}=1$ for all $N$.\footnote{More generally, we can consider free chiral multiplets in a more general complex representation $\mathbf{R}$ of $SU(N)$.  The ratio $K/\tau^{3/2}$ is proportional to $A(\mathbf{R} ) /T(\mathbf{R})^{3/2}$ where $\Tr_\mathbf{R} (\{T^a,T^b\}T^c) =
A(\mathbf{R}) d^{abc}$ and $\Tr_\mathbf{R} (T^aT^b) = T(\mathbf{R})\delta^{ab} $ are the anomaly coefficient and the index of the representation $\mathbf{R}$, respectively.  We find that the fundamental representation always maximizes the ratio $|K|/\tau^{3/2}$.}

For SQCD with $SU(N_c)$ gauge group, there is an $SU(N_f)$ flavor symmetry that only acts on the $N_f$ $SU(N_c)$-fundamental chiral multiplets, but not on the $N_f$ $SU(N_c)$-antifundamental chiral multiplets. The 't Hooft anomaly of this $SU(N_f)$ can be computed at zero coupling, which is simply $K=N_c$.  In the UV, we can choose a $U(1)_R$  symmetry under which both the fundamental and the anti-fundamental squarks have charge ${(N_f-N_c)/N_f}$.
In the UV QFT, the mixed 't Hooft anomaly between the $SU(N_f)$ flavor symmetry and the $R$-symmetry is $\Tr(R T^aT^b ) = - {N_c^2\over N_f}\delta^{ab}$.
Inside the conformal window ${3N_c/2} < N_f < 3N_c$ the UV QFT flows to an interacting IR SCFT \cite{Seiberg:1994pq} and the UV $R$-symmetry becomes the $R$-symmetry of the IR SCFT. The mixed 't Hooft anomaly then immediately gives the flavor current two-point function in the IR SCFT, $\tau \delta^{ab} = -3\Tr(R T^aT^b ) = {3N_c^2\over N_f} \delta^{ab}$ \cite{Anselmi:1997am,Anselmi:1997ys}.\footnote{The $U(1)_B$ baryon number symmetry does not commute with charge conjugation, and is therefore forbidden to mix with the $R$ symmetry \cite{Seiberg:1994pq,Intriligator:2003jj}. However, outside the conformal window, when the theory is IR free, other $U(1)$ symmetries can mix, and the UV $U(1)_R$ symmetry is generally not the IR $R$-symmetry.  For example, when $N_f>3N_c$, the theory is IR free, and the IR $R$-symmetry is a mixture of the UV $U(1)_R$ with the axial symmetry (the latter is free of the ABJ anomaly in the deep IR when the gauge coupling is off).  Therefore, the formula $\tau = {3N_c^2\over N_f}$ is not applicable outside the conformal window, except possibly at the boundary.
}
Hence, the ratio for the $SU(N_f)$ flavor symmetry in  the $SU(N_c)$ SQCD is
\ie
SU(N_c)~\text{SQCD}:~~{K\over \tau^{3/2}  } =  {1\over 3^{3/2} } {N_f^{3/2} \over N_c^2}\,. \label{eq:SQCD_SUN_Ratio}
\fe
Similarly for SQCD with $SO(N_c)$ gauge group ($N_c\ge 3$), there is an $SU(N_f)$ flavor symmetry rotating the $N_f$  chiral multiplets in the vector representations of $SO(N_c)$.  The $SU(N_f)$ flavor symmetry has the 't Hooft anomaly $K=N_c$.  In the UV, we define the $U(1)_R$ symmetry so that the squark has charge $(N_f-(N_c-2) )/N_f$.  In the conformal window $3(N_c-2)/2 < N_f < 3(N_c-2)$, the UV $U(1)_R$ symmetry is identified with the IR $R$-symmetry and  the two-point function of the flavor current in the IR is given by the mixed flavor-$R$ anomaly to be $\tau = {3 N_c(N_c-2)\over N_f}$. Hence, the ratio for the $SU(N_f)$ flavor symmetry in the $SO(N_c)$ SQCD is
\ie
SO(N_c)~\text{SQCD}:~~{K\over \tau^{3/2} }  = { 1 \over 3^{3/2}  }  { N_f^{3/2} \over N_c^{1/2} (N_c-2)^{3/2} }\,.  \label{eq:SQCD_SON_Ratio}
\fe

\section{Bootstrapping the Flavor Currents}
\label{sec:Bootstrap}

\subsec{Outline of the method}[secBootOutline]
Let us begin this section by illustrating how $\tau$ and $K/\tau^{3/2}$ appear in OPE coefficients of the $J^a\times J^b$ OPE, and the logic behind the method we will be using to obtain our OPE coefficient bounds.

What we will be bounding below are squares of OPE coefficients in the $J^a\times J^b$ OPE. To figure out their relation to physical quantities like $\tau$ and $K/\tau^{3/2}$, we need to carefully keep track of all normalizations and recall that in the bootstrap we use unit-normalized operators. The unit-normalized operators $\hat{J}^a$ and $\hat{\jmath}^{\lsp a}_\mu$ of interest here are related to $J^a$ and $j^a_\mu$ by $J^a=\sqrt{\frac{\tau}{4}}\frac{1}{2\pi^2}\hat{J}^{a}$ and $j^a_\mu=\frac{\sqrt{3\tau}}{2\pi^2}\hat{\jmath}_\mu^{\lsp a}\,.$
Starting from \eqref{JJOPE}, we have
\eqn{
\langle J^a J^b J^c\rangle\sim\frac{K}{64\pi^6}d^{\llsp abc} \quad\Rightarrow\quad \langle \hat{J}^a \hat{J}^b \hat{J}^c\rangle
\sim\frac{K}{\tau^{3/2}}d^{\llsp abc}\,,
}[]
where we only write down the dependence on the overall coefficient.  Dropping the hats, the OPE coefficient of $J^c$ in the $J^a\times J^b$ OPE is given by
\eqn{\lambda_{JJJ}=\frac{K}{\tau^{3/2}}\,.}[OPECoeffJJJ]
This is true for abelian and nonabelian symmetries.

Similarly, equations \eqref{JJOPE} and the nonabelian version of \eqref{jjform} imply that
\eqn{\langle J^a J^b j_\mu^c\rangle\sim\frac{\tau}{32\pi^6}f^{abc}
  \quad\Rightarrow\quad \langle \hat{J}^a \hat{J}^b
\hat{j}_\mu^c\rangle\sim\frac{1}{\sqrt{3\tau}}f^{abc}\,.}[]
Dropping the hats again we see that the OPE coefficient of the $j_\mu^c$ operator in the $J^a\times J^b$ OPE is
\eqn{\lambda_{JJj_\mu}=\frac{1}{\sqrt{3\tau}}\,.}[OPECoeffJJj]
This only appears in the nonabelian case.

As we will see below, the bootstrap allows us to bound squares of OPE coefficients, particularly $\lambda_{JJJ}^2$ and $\lambda_{JJj_\mu}^2$.  These can obviously be translated to bounds on $|\lambda_{JJJ}|=|K|/\tau^{3/2}$ and $\lambda_{JJj_\mu}$.

To explain how to bound OPE coefficients like \OPECoeffJJJ and \OPECoeffJJj, consider a single four-point function of a scalar operator $\phi(x)$ in a nonsupersymmetric theory. This is the simplest situation in which we can illustrate the logic~\cite{Caracciolo:2009bx}.  The conformal block decomposition of this four-point function in the $12\rightarrow 34$ channel is
\eqn{\langle\phi(x_1)\phi(x_2)\phi(x_3)\phi(x_4)\rangle =
\frac{1}{x_{12}^{2\Delta_\phi}x_{34}^{2\Delta_\phi}}\sum_{\mathcal{O}}
\lambda_{\phi\phi\mathcal{O}}^2
\, g_{\mathcal{O}}(u,v)\,,}[confBlockI]
where $\lambda_{\phi\phi\mathcal{O}}$ is the OPE coefficient of $\mathcal{O}$ in the $\phi\times\phi$ OPE and $g_{\mathcal{O}}(u,v)$ is the conformal block that depends on the dimension $\Delta$ and spin $\ell$ of the operator $\mathcal{O}$ as well as the conformally invariant cross ratios
\eqn{u=\frac{x_{12}^{2}x_{34}^{2}}{x_{13}^{2}x_{24}^{2}}\,,\qquad
v=\frac{x_{14}^{2}x_{23}^{2}}{x_{13}^{2}x_{24}^{2}}\,.}[]
Another conformal-block decomposition can be obtained by considering the $14\rightarrow 32$ channel:
\eqn{\langle\phi(x_1)\phi(x_2)\phi(x_3)\phi(x_4)\rangle =
\frac{1}{x_{14}^{2\Delta_\phi}x_{23}^{2\Delta_\phi}}
\sum_{\mathcal{O}}\lambda_{\phi\phi\mathcal{O}}^2
\, g_{\mathcal{O}}(v,u)\,.}[confBlockII]
Equating the right-hand sides of \confBlockI and \confBlockII gives rise to the crossing equation:
\eqn{\sum_{\mathcal{O}}\lambda_{\phi\phi\mathcal{O}}^2 F_{\mathcal{O}}(u,v)=0\,,\qquad F_{\mathcal{O}}(u,v)=u^{-\Delta_\phi} g_{\mathcal{O}}(v,u)-v^{-\Delta_\phi}g_{\mathcal{O}}(u,v)\,.}[crEq]

In order to obtain a bound on the square of the OPE coefficient of a given operator $\mathcal{O}_0$ in the set of all $\mathcal{O}$'s, we write \crEq in the following way:
\eqn{\lambda_{\phi\phi\mathcal{O}_0}^2 F_{\mathcal{O}_0}(u,v)
= -F_{\mathds{1}}-\sum_{\mathcal{O}\neq\mathds{1},\mathcal{O}_0} \lambda_{\phi\phi\mathcal{O}}^2 F_{\mathcal{O}}(u,v)\,,}[crEqOPE]
where, as usual, we have normalized our operators $\phi$ such that the OPE coefficient of the identity operator $\mathds{1}$ is equal to 1. Now we act on \crEqOPE with a linear functional $\alpha$ and demand
\eqn{\alpha(F_{\mathcal{O}_0})=1\quad\text{and}\quad
\alpha(F_\mathcal{O})\geq 0\,, \text{ for all }\mathcal{O}
\text{'s in the sum}.}[funcConstr]
Then, assuming unitarity, which implies that all $\lambda_{\phi\phi\mathcal{O}}$'s are real, we obtain
\eqn{\lambda_{\phi\phi\mathcal{O}_0}^2=-\alpha(F_{\mathds{1}})-(\text{positive)}\leq-\alpha(F_{\mathds{1}})\,.}[]
In the space of functionals that satisfy \funcConstr we choose the one that minimizes $-\alpha(F_\mathds{1})$ in order to get the strongest upper bound (within that space of functionals) for $\lambda_{\phi\phi\mathcal{O}_0}^2$.

\subsection{\texorpdfstring{$U(1)$}{U(1)}}
In this section we will briefly review the conformal bootstrap equations for $U(1)$ flavor currents in 4$d$ $\mathcal{N}=1$ SCFTs before presenting the main results. Crossing symmetry for the four-point function $\langle J(x_1)J(x_2)J(x_3)J(x_4)\rangle$ implies \cite{Fortin:2011nq}
\ie
&v^{2}\left(\sum\limits_{\mathcal{O}_{\ell}}|c_{JJ\mathcal{O}_{\ell}}|^{2}\mathcal{G}_{\Delta,\ell}(u,v)+\sum\limits_{(Q\mathcal{O})_{\ell}}|c_{JJ(Q\mathcal{O})_{\ell}}|^{2}g_{\Delta,\ell}(u,v)\right)
\\
&\qquad=u^{2}\left(\sum\limits_{\mathcal{O}_{\ell}}|c_{JJ\mathcal{O}_{\ell}}|^{2}\mathcal{G}_{\Delta,\ell}(v,u)+\sum\limits_{(Q\mathcal{O})_{\ell}}|c_{JJ(Q\mathcal{O})_{\ell}}|^{2}g_{\Delta,\ell}(v,u)\right)\,,
\label{eq:sum4JAbelian}
\\
\\
&\mathcal{G}_{\Delta,\ell \hspace{.03in} \text{even}}=\
g_{\Delta,\ell}+\frac{(\Delta-2)^{2}(\Delta+\ell)(\Delta-\ell-2)}{16\Delta^{2}(\Delta-\ell-1)(\Delta+\ell+1)}g_{\Delta+2,\ell}\,,
\\
&\mathcal{G}_{\Delta,\ell \hspace{.03in}\text{odd}}= \
g_{\Delta+1,\ell+1}+\frac{(\ell+2)^{2}(\Delta+\ell+1)(\Delta-\ell-2)}{4\ell^{2}(\Delta+\ell)(\Delta-\ell-1)}g_{\Delta+1,\ell-1}\,.
\fe
The sum for $\mathcal{G}_{\Delta,\ell}$ in (\ref{eq:sum4JAbelian}) runs over superconformal primaries $\mathcal{O}_{\Delta,\ell}$ which appear in the $J\times J$ OPE. The second sum for $g_{\Delta, \ell}$ runs over conformal primary operators $(Q\mathcal{O})_{\Delta,\ell}$ that are superconformal descendents, which appear in the $J \times J$ OPE but whose superconformal primary, $\mathcal{O}$, does not. For these operators we have $\ell\geq2$ and even with $\Delta\geq\ell+3$~\cite{Berkooz:2014yda}.

To set the conventions we define the nonsupersymmetric blocks $g_{\Delta,\ell}$ and cross ratios as
\ie
&g_{\Delta,\ell}(z,\bar{z})=\Big(-\frac{1}{2}\Big)^\ell\frac{z\bar{z}}{z-\bar{z}}\lsp\big(k_{\Delta+\ell}(z)k_{\Delta-\ell-2}(z)-(z\leftrightarrow\bar{z})\big)\,,
\\
&k_{\beta}(z)=z^{\frac{\beta}{2}}{}_{2}F_{1}\big(\tfrac12\beta,\tfrac12\beta;
\beta;z\big)\,,
\\
&u=z\bar{z}, \quad
v=(1-z)(1-\bar{z})\,.
\fe
To derive constraints on the OPE data we follow the standard numerical
procedure reviewed in \cite{Poland:2018epd, Chester:2019wfx} and summarized in subsection \secBootOutline above. For numerical implementation we use \href{https://github.com/davidsd/sdpb/tree/elemental}{\texttt{SDPB}}~\cite{Simmons-Duffin:2015qma} and derivative functionals $\alpha=\sum\limits_{m,n}a_{mn}\partial^{m}_{z}\partial^{n}_{\bar{z}}$ with $m+n\leq\Lambda$. We can derive rigorous bounds at any finite $\Lambda$ and, by making an ansatz, make predictions for the limit $\Lambda\rightarrow \infty$.

\begin{figure}[H]
  \centering
\includegraphics[scale=.575]{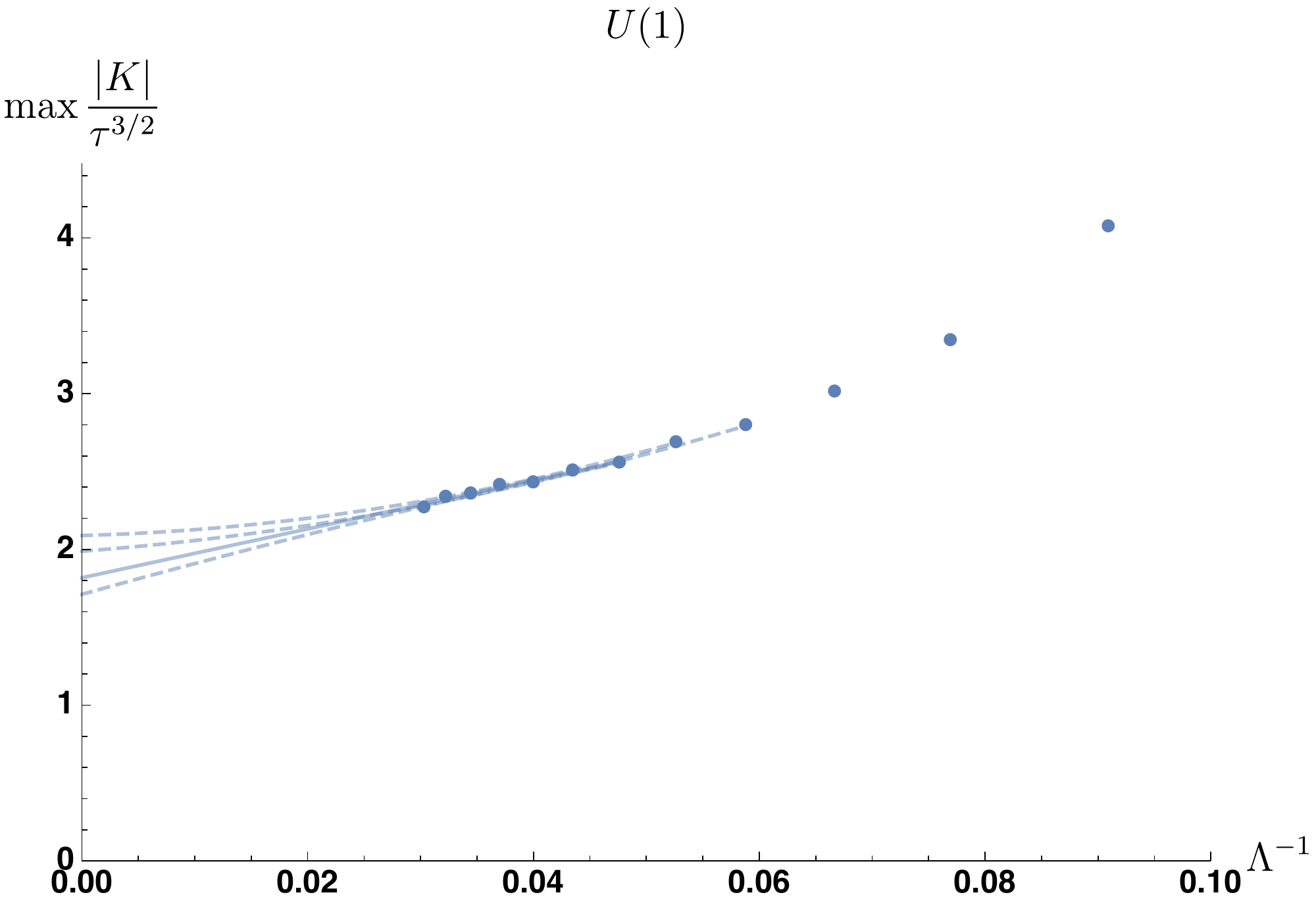}
  \caption{Upper bound on $|K|/\tau^{3/2}$ in theories with a $U(1)$
 flavor symmetry at different derivative orders $\Lambda$. The solid line is a linear fit extrapolated to an infinite number of derivatives, and the dashed lines are quadratic fits (in $\Lambda^{-1}$) with different data ranges.}
  \label{fig:U1_Bound_K}
\end{figure}

\begin{figure}[H]
  \centering
\includegraphics[scale=.575]{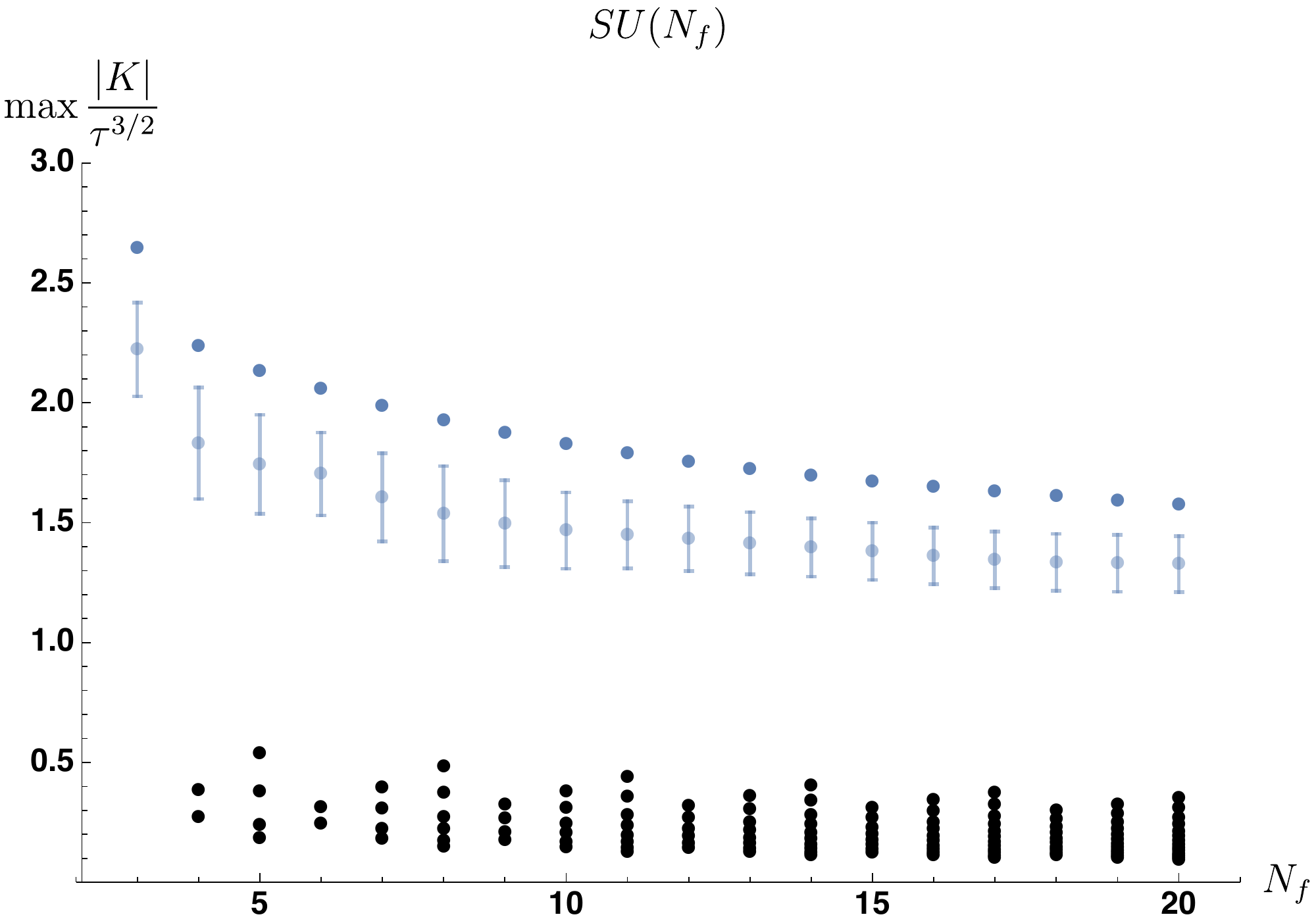}
  \caption{Upper bounds on $|K|/\tau^{3/2}$ in theories with $SU(N_f)$  flavor symmetry. The upper blue dots are rigorous upper bounds obtained at derivative order $\Lambda = 29$, while the blue error bars are upper bounds from extrapolation to $\Lambda = \infty$. See the discussion around \eqref{U1Bounds} for the method of extrapolation and error estimates. The black dots are various SQCDs with $SU(N_c)$ and $SO(N_c)$ gauge groups in the conformal window.}
  \label{fig:maxk}
\end{figure}

In Fig.~\ref{fig:U1_Bound_K} we have plotted the upper bound on the $\langle JJJ\rangle$ OPE coefficient as a function of $\Lambda^{-1}$.  The best rigorous bound obtained at $\Lambda = 33$, and the estimate of the
limit $\Lambda\rightarrow \infty$ based on linear and quadratic (in $\Lambda^{-1}$) extrapolations are
\begin{align}\label{U1Bounds}
  \frac{|K|}{\tau^{3/2}}\leq2.3 \qquad (\Lambda = 33),
  \hspace{.5in}
  \frac{|K|}{\tau^{3/2}}\leq1.9(2) \qquad (\Lambda = \infty)\,.
\end{align}
The error is estimated by varying the range of $\Lambda$ we use to extrapolate, and computing the standard deviation of the resulting extrapolated values.

\subsection{\texorpdfstring{$SU(N)$}{SU(N)} and the conformal window of SQCD}
For theories with an $SU(N)$ global symmetry the analysis is essentially unchanged, except that we now have a system of crossing equations. The number of crossing equations is equal to the number of $SU(N)$ representations which appear in the tensor products of two adjoints. This problem was studied in \cite{Berkooz:2014yda} so we will refer the reader there for more details.\footnote{Note that we normalize our nonsupersymmetric blocks differently, $g^{(\text{here})}_{\Delta,\ell}=\left(-\frac{1}{2}\right)^{\ell}g^{(\text{there})}_{\Delta,\ell}$.}

One important difference with the abelian case is that now both $\langle JJJ\rangle$ and $\langle JJj_{\mu}\rangle$ are nonzero\cite{Fortin:2011nq}. As reviewed in Section~\ref{sec:PertAnomalies}, the two OPE coefficients are independent and we
can bound both  $|K|/\tau^{3/2}$ and $\tau^{-1/2}$.

\begin{figure}[H]
  \centering
\includegraphics[scale=.6]{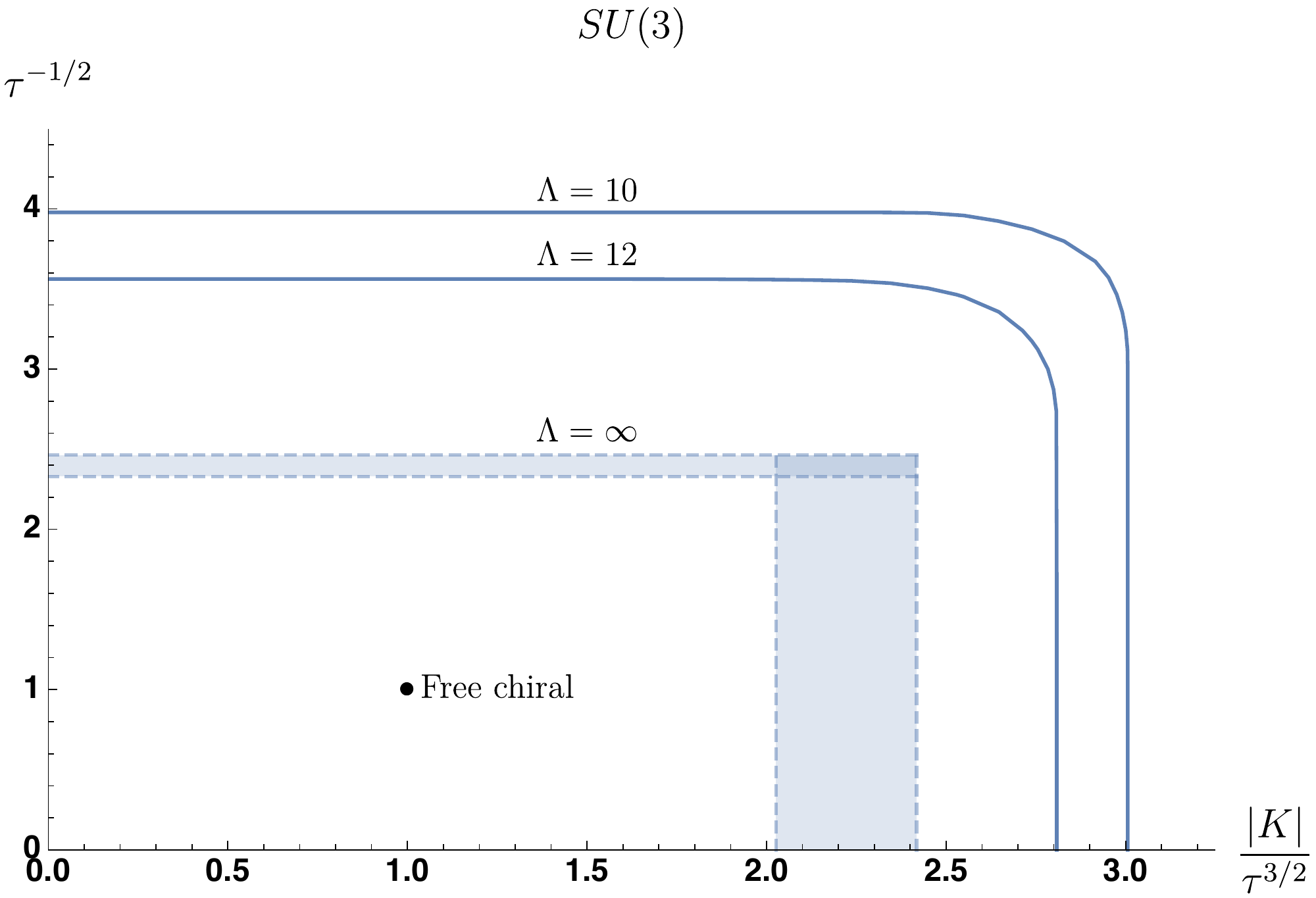}
  \caption{Allowed region in the $\tau$-$|K|$ plane for theories with an $SU(3)$ flavor symmetry, at derivative orders $\Lambda = 10, 12$.  The rectangle is obtained by extrapolations to $\Lambda = \infty$ of upper bounds on $\tau^{-1/2}$ and on $|K|/\tau^{3/2}$ (the actual allowed region is expected to be smaller than this rectangle). Also shown is the value of a free chiral multiplet in the fundamental representation of $SU(3)$. All other known theories have smaller $\tau^{-1/2}$ and smaller $|K|/\tau^{3/2}$ than the free chiral multiplet.}
  \label{fig:SU3}
\end{figure}

In Table~\ref{table:SQCD} and Figure~\ref{fig:maxk}, we present the universal upper bounds on $|K|/\tau^{3/2}$ across a range of $N_f$, and compare with SQCDs with $SU(N_c)$ and $SO(N_c)$ gauge groups in the conformal window predicted by Seiberg.
See the discussion around \eqref{U1Bounds} for the method of extrapolation and error estimates.
In Figure~\ref{fig:SU3} we consider theories with an $SU(3)$ global symmetry and exhibit the allowed region as a function of $\tau$ and $|K|$. For most of the parameter space the two bounds are independent, but as we approach the corner of the allowed region they appear to be correlated.  Since the universal upper bounds on $\tau^{-1/2}$ have already been studied, we present the improved bounds in Appendix~\ref{app:tau}.

\begin{table}[H]
\begin{centering}
	\begin{tabular}{|c|c|c|c|c|c|c|}
	\hline
	$N_f$ & $\Lambda = 29$ & $\Lambda = \infty$ & $SU(N_c)$ SQCD & $SO(N_c)$ SQCD
	\\ \hline
 3 & 2.6 & \text{2.2(2)} & &  \\
 4 & 2.2 & \text{1.8(2)} & 0.38 & 0.27 \\
 5 & 2.1 & \text{1.7(2)} & 0.54 & 0.38 \\
 6 & 2.1 & \text{1.7(2)} & 0.31 & 0.24 \\
 7 & 2.0 & \text{1.6(2)} & 0.40 & 0.31 \\
 8 & 1.9 & \text{1.5(2)} & 0.48 & 0.37 \\
 9 & 1.9 & \text{1.5(2)} & 0.32 & 0.27 \\
 10 & 1.8 & \text{1.5(2)} & 0.38 & 0.31 \\
 11 & 1.8 & \text{1.4(1)} & 0.44 & 0.36 \\
 12 & 1.8 & \text{1.4(1)} & 0.32 & 0.27 \\
 13 & 1.7 & \text{1.4(1)} & 0.36 & 0.30 \\
 14 & 1.7 & \text{1.4(1)} & 0.40 & 0.34 \\
 15 & 1.7 & \text{1.4(1)} & 0.31 & 0.27 \\
 16 & 1.6 & \text{1.4(1)} & 0.34 & 0.30 \\
 17 & 1.6 & \text{1.3(1)} & 0.37 & 0.32 \\
 18 & 1.6 & \text{1.3(1)} & 0.30 & 0.26 \\
 19 & 1.6 & \text{1.3(1)} & 0.33 & 0.29 \\
 20 & 1.6 & \text{1.3(1)} & 0.35 & 0.31 \\
	\hline
\end{tabular}
\caption{\label{table:SQCD}
The bootstrap upper bounds on the ratio $|K|/\tau^{3/2}$ for an $SU(N_f)$ flavor symmetry are shown in the second and the third columns.  The numerical results at derivative order $\Lambda = 29$ and from extrapolating to infinite derivative order are both included.  For a fixed $N_f$, the fourth and the fifth columns show the maximum values of the ratio $|K|/\tau^{3/2}$ for SQCD in the conformal window, which are all consistent with the bootstrap bounds.
}
\end{centering}
	\end{table}

\section{Discussion}
\label{sec:Discussion}

In this paper we use the conformal bootstrap method to derive universal constraints on 't Hooft anomalies in (3+1)$d$ superconformal field theories.  The existence of an upper bound on $K/\tau^{3/2}$ matches with our expectation that the anomaly is constrained by the amount of charged degrees of freedom in a theory, or simply, ``central charge $>$ anomaly".  Since the 't Hooft anomaly is invariant under renormalization group flows, we further apply this bound to compare with the conformal window of SQCD.

Another application of our result we have not discussed thus far is to AdS/CFT\cite{Maldacena:1997re,Witten:1998qj,Gubser:1998bc}. As pointed out originally in \cite{Witten:1998qj}, the 't Hooft anomaly of conserved currents in a (3+1)$d$ CFT is dual to a Chern--Simons term in AdS$_{5}$. More precisely, the AdS action for the gauge sector is \cite{Freedman:1998tz}:
\begin{align}
S=\int\limits_{AdS_{5}}d^{5}x\left[\sqrt{g}\frac{1}{4g^{2}}F^{a}_{\mu\nu}F^{\mu\nu
a}+\frac{ik}{96\pi^{2}}\left(d^{abc}\epsilon^{\mu\nu\lambda\rho\sigma}
A^{a}_{\mu}\partial_{\nu}A^{b}_{\lambda}\partial_{\rho}A^{c}_{\sigma}+\ldots\right)\right]
\end{align}
where the $``\ldots"$ is the completion of the $(4+1)d$ Chern--Simons term. Given our normalization, the relation between the CFT and AdS quantities is
\begin{align}
\tau = \frac{8\pi^{2}R_{AdS}}{g^{2}}, \qquad K=k \,,
\end{align}
so the upper bound on $|K|/\tau^{3/2}$ turns into an upper bound on $kg^{3}/R_{AdS}^{3/2}$. In other words, unitarity and locality of the dual CFT prevents us from making the bulk Chern--Simons term arbitrarily dominant over the Yang--Mills term.\footnote{An extreme limit that the bootstrap rules out is the pure (4+1)$d$ Chern--Simons theory without a Yang--Mills term. Indeed, it has been a longstanding question on how to quantize the pure (4+1)$d$ Chern--Simons theory consistently. We thank K.\ Ohmori for discussions.
}
The bound we derived here follows from basic CFT axioms, and it would also be interesting to understand if this bound can be strengthened for large $N$ CFTs with a large gap in the higher spin spectrum \textit{\`a la} \cite{Camanho:2014apa}.

There are several generalizations that can potentially improve the constraints on  SQCD in the conformal window.  One natural generalization would be to include the other conformal primaries in the flavor supermultiplet as external operators, such as the flavor current itself. As demonstrated in \cite{Cornagliotto:2017dup} for $\mathcal{N}=2$ in (1+1)$d$, including such operators significantly strengthens the bounds (see also \cite{Bissi:2015qoa,Buric:2019rms}).  In this paper, with the external operators chosen to be the superconformal primary $J$, the quantities $\tau$ and $K/\tau^{3/2}$ appear as two independent OPE coefficients, and we find that the constraints on them are only weakly dependent (the allowed region in Figure~\ref{fig:SU3} is close to being rectangular). When the other conformal primaries are included as external operators, $\tau$ and $K/\tau^{3/2}$ appear in mixed ways in the OPE coefficients, and we can hope for a more nontrivial profile for the allowed region.\footnote{We thank P.\ Kravchuk for a discussion on this point.
}

Since every (3+1)$d$ SCFT contains the $U(1)_{R}$ supercurrent multiplet, it would also be interesting to study mixed systems involving these operators. In particular, this would give us access to mixed anomalies between the flavor and $R$-symmetries. These problems all require understanding the (3+1)$d$ spinning (super-)conformal blocks, which we expect will reveal new information about the space of (3+1)$d$ SCFTs.

Finally, it would be interesting to explore further the implications of 't Hooft anomalies for the CFT data.  Does the triangle anomaly in (3+1)$d$ place an upper bound on the lightest charged local operator in a CFT?  Does the {\it global} anomaly for $SU(2)$ global symmetry (the Witten anomaly) \cite{Witten:1982fp} place a bound on the lightest $SU(2)$ half-integral spin local operator?  One can also generalize this discussion to higher dimensions.  For instance, the 't Hooft (square) anomaly of a continuous global symmetry in (5+1)$d$ appears in a specific structure in the four-point function of the spin-one conserved current, which is also accessible from the conformal bootstrap.

\ack{We would like to thank M.\ Barkeshli, P.\ Kravchuk, K.\ Ohmori, and
S.\ Razamat for enlightening discussions.  The computations in this paper
are performed on the Helios computing cluster supported by the School of
Natural Sciences Computing Staff at the Institute for Advanced Study.  This
research was supported in part by Perimeter Institute for Theoretical
Physics.  Research at Perimeter Institute is supported by the Government of
Canada through the Department of Innovation, Science and Economic
Development and by the Province of Ontario through the Ministry of Research
and Innovation.  This work was performed in part at Aspen Center for
Physics, which is supported by National Science Foundation grant
PHY-1607611. This work was partially supported by a grant from the Simons
Foundation.  The research of YL and DM is supported by the Walter Burke
Institute for Theoretical Physics and the Sherman Fairchild Foundation.
The work of  SHS is supported  by the National Science Foundation grant
PHY-1606531, the Roger Dashen Membership, and  a grant from the Simons
Foundation/SFARI (651444, NS).  Research presented in this article was
supported by the Laboratory Directed Research and Development program of
Los Alamos National Laboratory under project number 20180709PRD1. This
material is based upon work supported by the U.S. Department of Energy,
Office of Science, Office of High Energy Physics, under Award Number
DE-SC0011632.}

\begin{appendices}

\section{Bounds on \texorpdfstring{$\boldsymbol{\tau}$}{tau}}
\label{app:tau}

\begin{figure}[H]
  \centering
  \includegraphics[scale=.6]{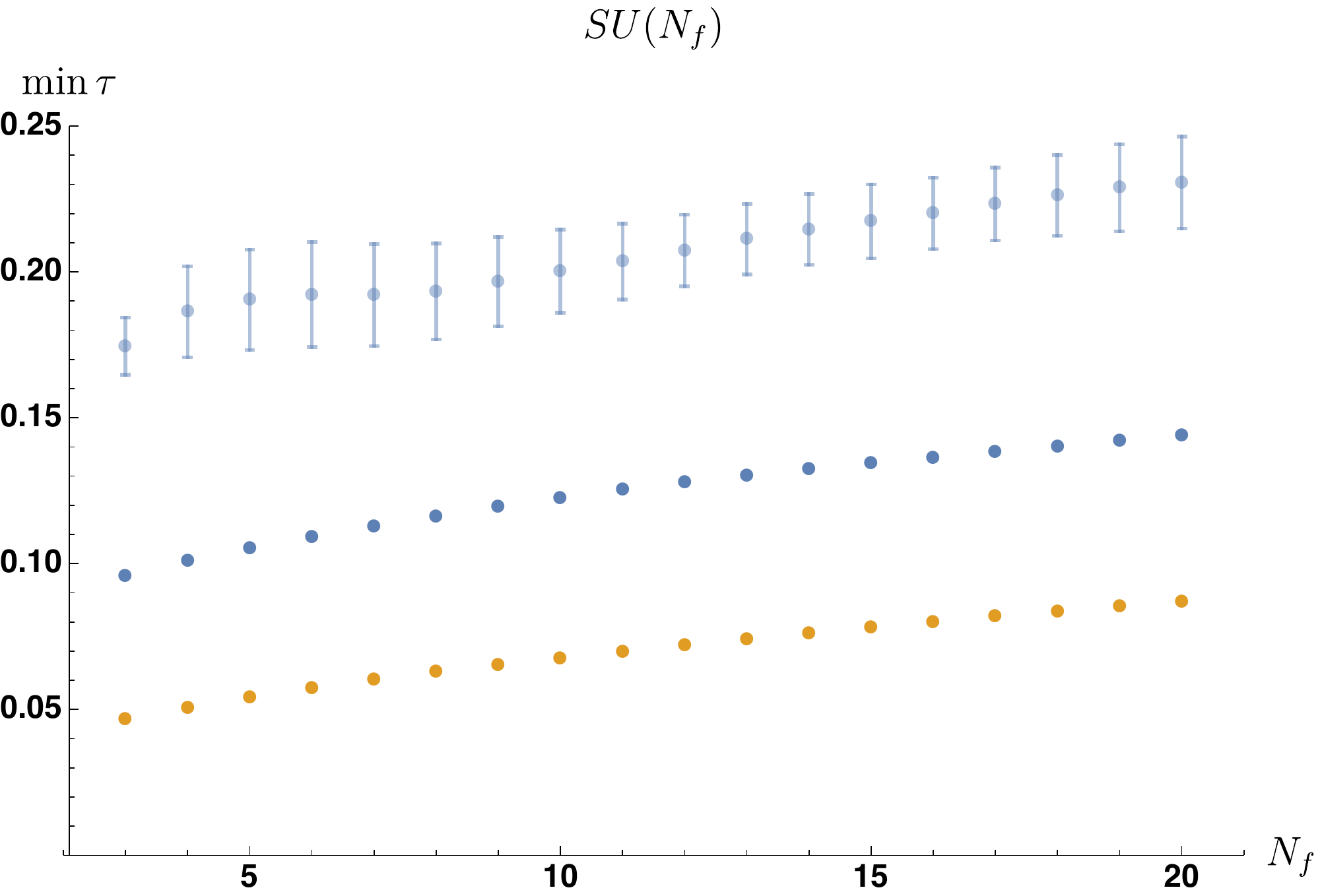}
  \caption{Lower bounds on $\tau$ in theories with $SU(N_f)$ flavor symmetry. The middle blue dots are rigorous lower bounds obtained at derivative order $\Lambda = 29$, while the blue error bars are lower bounds from extrapolation to $\Lambda = \infty$. For comparison, the bottom yellow dots are bounds at $\Lambda = 15$, which are similar to the bounds obtained by \cite{Berkooz:2014yda}.}
  \label{fig:mintau}
\end{figure}

The bootstrap setup of Section~\ref{sec:Bootstrap} also puts universal lower bounds on $\tau$, related to mixed flavor-$R$ 't Hooft anomaly in (3+1)$d$ superconformal field theory. Such bounds were first obtained in \cite{Poland:2010wg,Poland:2011ey} by bootstrapping the four-point function of general scalars, and in \cite{Berkooz:2014yda} by bootstrapping the four-point function of scalars in conserved current supermultiplets.
Then the $SU(3)$ case in the latter approach was studied in much further detail (with dependence on the gap) in \cite{Li:2017ddj}. In Figure~\ref{fig:mintau}, we present improved bounds on $\tau$ for general $SU(N_f)$ global symmetry under the same setup as in \cite{Berkooz:2014yda}, but performed with more advanced numerical technology up to derivative order $\Lambda = 29$ and with extrapolation to $\Lambda = \infty$. By contrast, the bounds of \cite{Berkooz:2014yda} are at $\Lambda = 15$, on which our $\Lambda = 29$ and $\Lambda = \infty$ bounds improve roughly by factors of 2 and 3, respectively.

\end{appendices}

\bibliography{Bound_Anomaly_4d_N=1}

\begin{thebibliography}{10}
\ifx\href\asklfhas\newcommand{\href}[2]{#2}\fi
\ifx\arxivref\asklfhas\newcommand{\arxivref}[2]{\href{http://arxiv.org/abs/#1}{#2}}\fi
\ifx\doiref\asklfhas\newcommand{\doiref}[2]{\href{http://dx.doi.org/#1}{#2}}\fi
\parskip 0pt
\normalsize

\bibitem{Seiberg:1994pq}
N.~Seiberg,
\textit{``{Electric-magnetic duality in supersymmetric nonAbelian gauge
  theories}''},
\doiref{10.1016/0550-3213(94)00023-8}{Nucl.~Phys. \textbf{B435}, 129
  (1995)\ignorespaces}\ignorespaces,
\normalsize{\texttt{\arxivref{hep-th/9411149}{hep-th/9411149}}}\ignorespaces
\bibitem{Rattazzi:2008pe}
R.~Rattazzi, V.~S. Rychkov, E.~Tonni \& A.~Vichi,
\textit{``{Bounding scalar operator dimensions in 4D CFT}''},
\doiref{10.1088/1126-6708/2008/12/031}{JHEP \textbf{0812}, 031
  (2008)\ignorespaces}\ignorespaces,
\normalsize{\texttt{\arxivref{0807.0004}{arXiv:0807.0004
  \![hep-th]}}}\ignorespaces
\bibitem{Poland:2018epd}
D.~Poland, S.~Rychkov \& A.~Vichi,
\textit{``{The Conformal Bootstrap: Theory, Numerical Techniques, and
  Applications}''},
\doiref{10.1103/RevModPhys.91.015002}{Rev.~Mod.~Phys. \textbf{91}, 015002
  (2019)\ignorespaces}\ignorespaces,
\normalsize{\texttt{\arxivref{1805.04405}{arXiv:1805.04405
  \![hep-th]}}}\ignorespaces
\bibitem{Chester:2019wfx}
S.~M. Chester,
\textit{``{Weizmann Lectures on the Numerical Conformal Bootstrap}''},
\normalsize{\texttt{\arxivref{1907.05147}{arXiv:1907.05147
  \![hep-th]}}}\ignorespaces
\bibitem{ElShowk:2012ht}
S.~El-Showk, M.~F. Paulos, D.~Poland, S.~Rychkov, D.~Simmons-Duffin \&
  A.~Vichi,
\textit{``{Solving the 3D Ising Model with the Conformal Bootstrap}''},
\doiref{10.1103/PhysRevD.86.025022}{Phys.~Rev. \textbf{D86}, 025022
  (2012)\ignorespaces}\ignorespaces,
\normalsize{\texttt{\arxivref{1203.6064}{arXiv:1203.6064
  \![hep-th]}}}\ignorespaces
\bibitem{El-Showk:2014dwa}
S.~El-Showk, M.~F. Paulos, D.~Poland, S.~Rychkov, D.~Simmons-Duffin \&
  A.~Vichi,
\textit{``{Solving the 3d Ising Model with the Conformal Bootstrap II.
  c-Minimization and Precise Critical Exponents}''},
\doiref{10.1007/s10955-014-1042-7}{J.~Stat.~Phys. \textbf{157}, 869
  (2014)\ignorespaces}\ignorespaces,
\normalsize{\texttt{\arxivref{1403.4545}{arXiv:1403.4545
  \![hep-th]}}}\ignorespaces
\bibitem{Kos:2014bka}
F.~Kos, D.~Poland \& D.~Simmons-Duffin,
\textit{``{Bootstrapping Mixed Correlators in the 3D Ising Model}''},
\doiref{10.1007/JHEP11(2014)109}{JHEP \textbf{1411}, 109
  (2014)\ignorespaces}\ignorespaces,
\normalsize{\texttt{\arxivref{1406.4858}{arXiv:1406.4858
  \![hep-th]}}}\ignorespaces
\bibitem{Kos:2016ysd}
F.~Kos, D.~Poland, D.~Simmons-Duffin \& A.~Vichi,
\textit{``{Precision Islands in the Ising and $O(N)$ Models}''},
\doiref{10.1007/JHEP08(2016)036}{JHEP \textbf{1608}, 036
  (2016)\ignorespaces}\ignorespaces,
\normalsize{\texttt{\arxivref{1603.04436}{arXiv:1603.04436
  \![hep-th]}}}\ignorespaces
\bibitem{Lin:2019kpn}
Y.-H. Lin \& S.-H. Shao,
\textit{``{Anomalies and Bounds on Charged Operators}''},
\doiref{10.1103/PhysRevD.100.025013}{Phys.~Rev. \textbf{D100}, 025013
  (2019)\ignorespaces}\ignorespaces,
\normalsize{\texttt{\arxivref{1904.04833}{arXiv:1904.04833
  \![hep-th]}}}\ignorespaces
\bibitem{Adler:1969gk}
S.~L. Adler,
\textit{``{Axial vector vertex in spinor electrodynamics}''},
\doiref{10.1103/PhysRev.177.2426}{Phys.~Rev. \textbf{177}, 2426
  (1969)\ignorespaces}\ignorespaces
\bibitem{Bell:1969ts}
J.~S. Bell \& R.~Jackiw,
\textit{``{A PCAC puzzle: $\pi^0 \to \gamma \gamma$ in the $\sigma$ model}''},
\doiref{10.1007/BF02823296}{Nuovo~Cim. \textbf{A60}, 47
  (1969)\ignorespaces}\ignorespaces
\bibitem{Anselmi:1997am}
D.~Anselmi, D.~Z. Freedman, M.~T. Grisaru \& A.~A. Johansen,
\textit{``{Nonperturbative formulas for central functions of supersymmetric
  gauge theories}''},
\doiref{10.1016/S0550-3213(98)00278-8}{Nucl.~Phys. \textbf{B526}, 543
  (1998)\ignorespaces}\ignorespaces,
\normalsize{\texttt{\arxivref{hep-th/9708042}{hep-th/9708042}}}\ignorespaces
\bibitem{Dymarsky:2017xzb}
A.~Dymarsky, J.~Penedones, E.~Trevisani \& A.~Vichi,
\textit{``{Charting the space of 3D CFTs with a continuous global symmetry}''},
\doiref{10.1007/JHEP05(2019)098}{JHEP \textbf{1905}, 098
  (2019)\ignorespaces}\ignorespaces,
\normalsize{\texttt{\arxivref{1705.04278}{arXiv:1705.04278
  \![hep-th]}}}\ignorespaces
\bibitem{Dymarsky:2017yzx}
A.~Dymarsky, F.~Kos, P.~Kravchuk, D.~Poland \& D.~Simmons-Duffin,
\textit{``{The 3d Stress-Tensor Bootstrap}''},
\doiref{10.1007/JHEP02(2018)164}{JHEP \textbf{1802}, 164
  (2018)\ignorespaces}\ignorespaces,
\normalsize{\texttt{\arxivref{1708.05718}{arXiv:1708.05718
  \![hep-th]}}}\ignorespaces
\bibitem{Cuomo:2017wme}
G.~F. Cuomo, D.~Karateev \& P.~Kravchuk,
\textit{``{General Bootstrap Equations in 4D CFTs}''},
\doiref{10.1007/JHEP01(2018)130}{JHEP \textbf{1801}, 130
  (2018)\ignorespaces}\ignorespaces,
\normalsize{\texttt{\arxivref{1705.05401}{arXiv:1705.05401
  \![hep-th]}}}\ignorespaces
\bibitem{Karateev:2019pvw}
D.~Karateev, P.~Kravchuk, M.~Serone \& A.~Vichi,
\textit{``{Fermion Conformal Bootstrap in 4d}''},
\doiref{10.1007/JHEP06(2019)088}{JHEP \textbf{1906}, 088
  (2019)\ignorespaces}\ignorespaces,
\normalsize{\texttt{\arxivref{1902.05969}{arXiv:1902.05969
  \![hep-th]}}}\ignorespaces
\bibitem{KKMSVprep}
D.~Karateev, P.~Kravchuk, A.~Manenti, A.~Stergiou \& A.~Vichi,
In progress\ignorespaces
\bibitem{Poland:2010wg}
D.~Poland \& D.~Simmons-Duffin,
\textit{``{Bounds on 4D Conformal and Superconformal Field Theories}''},
\doiref{10.1007/JHEP05(2011)017}{JHEP \textbf{1105}, 017
  (2011)\ignorespaces}\ignorespaces,
\normalsize{\texttt{\arxivref{1009.2087}{arXiv:1009.2087
  \![hep-th]}}}\ignorespaces
\bibitem{Poland:2011ey}
D.~Poland, D.~Simmons-Duffin \& A.~Vichi,
\textit{``{Carving Out the Space of 4D CFTs}''},
\doiref{10.1007/JHEP05(2012)110}{JHEP \textbf{1205}, 110
  (2012)\ignorespaces}\ignorespaces,
\normalsize{\texttt{\arxivref{1109.5176}{arXiv:1109.5176
  \![hep-th]}}}\ignorespaces
\bibitem{Berkooz:2014yda}
M.~Berkooz, R.~Yacoby \& A.~Zait,
\textit{``{Bounds on $\mathcal{N} = 1$ superconformal theories with global
  symmetries}''},
\doiref{10.1007/JHEP01(2015)132, 10.1007/JHEP08(2014)008}{JHEP \textbf{1408},
  008 (2014)\ignorespaces}\ignorespaces,
\normalsize{\texttt{\arxivref{1402.6068}{arXiv:1402.6068
  \![hep-th]}}}\ignorespaces,
[Erratum: JHEP01,132(2015)]\ignorespaces
\bibitem{Li:2017ddj}
D.~Li, D.~Meltzer \& A.~Stergiou,
\textit{``{Bootstrapping mixed correlators in 4D $ \mathcal{N} $ = 1 SCFTs}''},
\doiref{10.1007/JHEP07(2017)029}{JHEP \textbf{1707}, 029
  (2017)\ignorespaces}\ignorespaces,
\normalsize{\texttt{\arxivref{1702.00404}{arXiv:1702.00404
  \![hep-th]}}}\ignorespaces
\bibitem{Schreier:1971um}
E.~J. Schreier,
\textit{``{Conformal symmetry and three-point functions}''},
\doiref{10.1103/PhysRevD.3.980}{Phys.~Rev. \textbf{D3}, 980
  (1971)\ignorespaces}\ignorespaces
\bibitem{Fortin:2011nq}
J.-F. Fortin, K.~Intriligator \& A.~Stergiou,
\textit{``{Current OPEs in Superconformal Theories}''},
\doiref{10.1007/JHEP09(2011)071}{JHEP \textbf{1109}, 071
  (2011)\ignorespaces}\ignorespaces,
\normalsize{\texttt{\arxivref{1107.1721}{arXiv:1107.1721
  \![hep-th]}}}\ignorespaces
\bibitem{Fitzpatrick:2014oza}
A.~L. Fitzpatrick, J.~Kaplan, Z.~U. Khandker, D.~Li, D.~Poland \&
  D.~Simmons-Duffin,
\textit{``{Covariant Approaches to Superconformal Blocks}''},
\doiref{10.1007/JHEP08(2014)129}{JHEP \textbf{1408}, 129
  (2014)\ignorespaces}\ignorespaces,
\normalsize{\texttt{\arxivref{1402.1167}{arXiv:1402.1167
  \![hep-th]}}}\ignorespaces
\bibitem{Khandker:2014mpa}
Z.~U. Khandker, D.~Li, D.~Poland \& D.~Simmons-Duffin,
\textit{``{$ \mathcal{N} $ = 1 superconformal blocks for general scalar
  operators}''},
\doiref{10.1007/JHEP08(2014)049}{JHEP \textbf{1408}, 049
  (2014)\ignorespaces}\ignorespaces,
\normalsize{\texttt{\arxivref{1404.5300}{arXiv:1404.5300
  \![hep-th]}}}\ignorespaces
\bibitem{Anselmi:1997ys}
D.~Anselmi, J.~Erlich, D.~Z. Freedman \& A.~A. Johansen,
\textit{``{Positivity constraints on anomalies in supersymmetric gauge
  theories}''},
\doiref{10.1103/PhysRevD.57.7570}{Phys.~Rev. \textbf{D57}, 7570
  (1998)\ignorespaces}\ignorespaces,
\normalsize{\texttt{\arxivref{hep-th/9711035}{hep-th/9711035}}}\ignorespaces
\bibitem{Intriligator:2003jj}
K.~A. Intriligator \& B.~Wecht,
\textit{``{The Exact superconformal R symmetry maximizes a}''},
\doiref{10.1016/S0550-3213(03)00459-0}{Nucl.~Phys. \textbf{B667}, 183
  (2003)\ignorespaces}\ignorespaces,
\normalsize{\texttt{\arxivref{hep-th/0304128}{hep-th/0304128}}}\ignorespaces
\bibitem{Caracciolo:2009bx}
F.~Caracciolo \& V.~S. Rychkov,
\textit{``{Rigorous Limits on the Interaction Strength in Quantum Field
  Theory}''},
\doiref{10.1103/PhysRevD.81.085037}{Phys.~Rev. \textbf{D81}, 085037
  (2010)\ignorespaces}\ignorespaces,
\normalsize{\texttt{\arxivref{0912.2726}{arXiv:0912.2726
  \![hep-th]}}}\ignorespaces
\bibitem{Simmons-Duffin:2015qma}
D.~Simmons-Duffin,
\textit{``{A Semidefinite Program Solver for the Conformal Bootstrap}''},
\doiref{10.1007/JHEP06(2015)174}{JHEP \textbf{1506}, 174
  (2015)\ignorespaces}\ignorespaces,
\normalsize{\texttt{\arxivref{1502.02033}{arXiv:1502.02033
  \![hep-th]}}}\ignorespaces
\bibitem{Maldacena:1997re}
J.~M. Maldacena,
\textit{``{The Large N limit of superconformal field theories and
  supergravity}''},
Adv.Theor.Math.Phys. \textbf{2}, 231 (1998)\ignorespaces\ignorespaces,
\normalsize{\texttt{\arxivref{hep-th/9711200}{hep-th/9711200}}}\ignorespaces
\bibitem{Witten:1998qj}
E.~Witten,
\textit{``{Anti-de Sitter space and holography}''},
Adv.Theor.Math.Phys. \textbf{2}, 253 (1998)\ignorespaces\ignorespaces,
\normalsize{\texttt{\arxivref{hep-th/9802150}{hep-th/9802150}}}\ignorespaces
\bibitem{Gubser:1998bc}
S.~Gubser, I.~R. Klebanov \& A.~M. Polyakov,
\textit{``{Gauge theory correlators from noncritical string theory}''},
\doiref{10.1016/S0370-2693(98)00377-3}{Phys.Lett. \textbf{B428}, 105
  (1998)\ignorespaces}\ignorespaces,
\normalsize{\texttt{\arxivref{hep-th/9802109}{hep-th/9802109}}}\ignorespaces
\bibitem{Freedman:1998tz}
D.~Z. Freedman, S.~D. Mathur, A.~Matusis \& L.~Rastelli,
\textit{``{Correlation functions in the CFT(d) / AdS(d+1) correspondence}''},
\doiref{10.1016/S0550-3213(99)00053-X}{Nucl.~Phys. \textbf{B546}, 96
  (1999)\ignorespaces}\ignorespaces,
\normalsize{\texttt{\arxivref{hep-th/9804058}{hep-th/9804058}}}\ignorespaces
\bibitem{Camanho:2014apa}
X.~O. Camanho, J.~D. Edelstein, J.~Maldacena \& A.~Zhiboedov,
\textit{``{Causality Constraints on Corrections to the Graviton Three-Point
  Coupling}''},
\doiref{10.1007/JHEP02(2016)020}{JHEP \textbf{1602}, 020
  (2016)\ignorespaces}\ignorespaces,
\normalsize{\texttt{\arxivref{1407.5597}{arXiv:1407.5597
  \![hep-th]}}}\ignorespaces
\bibitem{Cornagliotto:2017dup}
M.~Cornagliotto, M.~Lemos \& V.~Schomerus,
\textit{``{Long Multiplet Bootstrap}''},
\doiref{10.1007/JHEP10(2017)119}{JHEP \textbf{1710}, 119
  (2017)\ignorespaces}\ignorespaces,
\normalsize{\texttt{\arxivref{1702.05101}{arXiv:1702.05101
  \![hep-th]}}}\ignorespaces
\bibitem{Bissi:2015qoa}
A.~Bissi \& T.~Lukowski,
\textit{``{Revisiting $ \mathcal{N}=4 $ superconformal blocks}''},
\doiref{10.1007/JHEP02(2016)115}{JHEP \textbf{1602}, 115
  (2016)\ignorespaces}\ignorespaces,
\normalsize{\texttt{\arxivref{1508.02391}{arXiv:1508.02391
  \![hep-th]}}}\ignorespaces
\bibitem{Buric:2019rms}
I.~Buric, V.~Schomerus \& E.~Sobko,
\textit{``{Superconformal Blocks: General Theory}''},
\normalsize{\texttt{\arxivref{1904.04852}{arXiv:1904.04852
  \![hep-th]}}}\ignorespaces
\bibitem{Witten:1982fp}
E.~Witten,
\textit{``{An SU(2) Anomaly}''},
\doiref{10.1016/0370-2693(82)90728-6}{Phys.~Lett. \textbf{117B}, 324
  (1982)\ignorespaces}\ignorespaces
\end{thebibliography}

\end{document}